# Proposal for a spin MOSFET based on spin gapless semiconductors

P. Graziosi

CNR – ISMN, via Gobetti 101, 40129, Bologna

patrizio.graziosi@gmail.com

*We propose a spintronic metal–oxide–semiconductor field effect transistor (spin MOSFET) where a spin gapless semiconductor (SGS) constitutes the channel and the drain is a ferromagnetic metal. SGS exhibit a non-zero band gap in only one of the spin sub-bands and feature complete spin polarization at finite temperature. We present an analytical model of the device and comment the properties relevant for devices applications. Our results boost SGS as a new paradigm for the spin MOSFET concept.*

The International Technology Roadmap for Semiconductors (ITRS) establishes the criteria for the next generation of spin polarized materials in the electronics industry:[1] **(i)** $T_C > 400$ K, **(ii)** carrier mediated magnetism, **(iii)** high spin polarization at room temperature, **(iv)** relatively high resistivity to avoid the conductivity mismatch[2,3] with semiconductors. ITRS comments on the failure of the most studied new magnetic materials in at least one of its criteria. At the other hand, spintronic field effect transistor (spin MOSFET) is regarded as promising possible low power devices enabling both information storage and reconfigurable logics.[1,4,5] Spin MOSFET uses spin polarized ferromagnetic source and drain (S and D) and their output depends on the relative orientation of S and D magnetization.[5,6] Thus, the search for novel spin polarized materials able to satisfy ITRS requirements is a timely activity.

The recently introduced family of spin gapless semiconductors (SGS)[7] promises to fulfill all the ITRS requests. SGS exhibit a non-zero band gap in one of the spin sub-bands and a zero band-gap for opposite spin orientation. In a standard gapless semiconductor, there is no gap between the occupied valence and the empty conduction band: the excitation of electrons from the valence band to the conduction band requires no energy. Noteworthy, in *spin* gapless semiconductors this excitation is activated for only one spin sub-band, leading to the 100% spin polarization and to the growth of magnetization with temperature in ferromagnetic samples.[8–11] The gapless condition can be realized between sub-bands with the same or opposite spin polarization, leading to conceptually different properties. They can feature full spin polarization of both electrons and holes or only one of the twos and the Fermi level position determines the sign and/or the amount of the spin polarization.[7] These materials could be either ferromagnetic, antiferromagnetic or para-/dia-magnetic, but, noteworthy, they always feature full spin polarization for at least one type of carriers. Hence, SGS can open a new playground in the usual ferromagnetic spintronics but also in the upcoming antiferromagnetic spintronics, where the absence of ferromagnetic elements enables enhanced downscaling.[12]

Two fundamental classes of materials present compounds with signatures of SGS behavior: doped $PbPdO_2$ and Heusler alloys. Table I summarizes their main electric and magnetic properties. It is possible to note that they fulfill all the ITRS criteria for next generation spin polarized materials. Figure 1a sketches the band structure that we assume, as suggested from both theoretical calculations and experiments on the doped $PbPdO_2$.[7,8] Such a band structure allows for spin-up electrons in the conduction band and spin up holes in the valence band.[13] The employment of SGS as S and D electrodes in spin-MOSFET should allow efficient spin injection (thanks to the matching of their conductivities), extremely high magnetoconductance ratio and competitive current drivability and subthreshold swing.[6]

**Table I**. Electric and magnetic properties of SGS materials. n is the carriers density, µ their mobility, $M_S$ stands for the saturation magnetization and $T_C$ for the Curie temperature in ferromagnetic samples; n.a. stands for not available.

| Material | Carrier type | n [cm$^{-3}$] | µ [V/(cm·s)] | Magnetic features | Ref. |
|---|---|---|---|---|---|
| $Mn_2CoAl$ | electrons | $2.2 \cdot 10^{17}$ | $6.8 \cdot 10^4$ | $T_C \approx 720$ K | 31 |
| $Mn_2CoAl$ | both | $4 \cdot 10^{22}$ | 0.56 | $T_C > 600$ K | 32 |
| $Mn_2CoAl$ | electrons | $1.6 \cdot 10^{20}$ | 0.45 | $T_C \approx 550$ K | 33 |
| $Ti_2MnAl$ | n.a. | n.a. | n.a. | $T_C > 650$ K, $M_S$ increasing with T | 11 |
| $PbPd_{0.9}Co_{0.1}O_2$ | Holes | $5 \cdot 10^{18}$ | 8.8 | paramagnetic | 27 |
| $PbPd_{0.81}Co_{0.19}O_2$ | n.a. | n.a. | n.a. | $T_C > 400$ K, $M_S$ increasing with T | 8 |

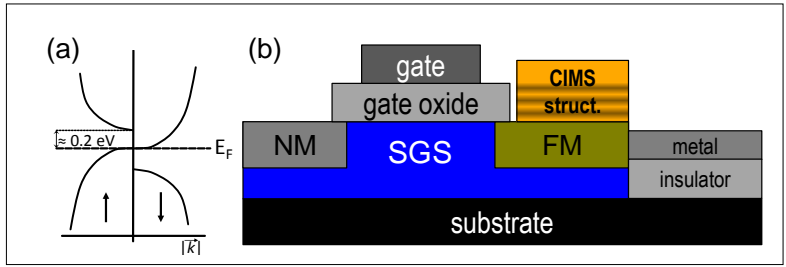

**Figure 1**: (a) sketch of the SGS energy bands adopted in this work. The gap in the gapped directions is about 0.2 eV, from calcuations;[7] (b) top gate geometry of SGS based spin-MOSFET, the arrows represents the relative spin orientations, NM stands for the non-magnetic metal S and FM for the ferromagnetic metal D. The CIMS structure is the multilayer structure for CIMS in FM,[14] which becomes part of the structure itself, and the right side metal closes the CIMS circuit.

In this letter, we show that the SGS band structure enables also a new kind of device, where the SGS is the channel material and only one of the S/D is a ferromagnetic metal. Figure 1a sketches the band structure, we assume that the thermally induced transition to the minority spin sub-band (with an Arrhenius probability) is the only factor that disturbs the spin polarization. We estimate a spin polarization value of 99.9% at room temperature. Figure 1b represents the proposed spin MOSFET structure in a top gate geometry; the bottom gate geometry will be suitable as well. The SGS is the channel material, the S contact is a normal metal (NM) such as Cu, Al, etc. and the D contact is a ferromagnetic metal (FM) like Ni, Fe, Co or Permalloy. The FM magnetization direction can be set by Current-Induced Magnetization Switch (CIMS)[5,14] by means of an *ad-hoc* CIMS structure (the FM is part of this CIMS structure) and a fourth contact is added to make the current to flow across the CIMS avoiding the FET for setting the FM magnetization direction. This allows for the memory option and hence reconfigurable logics.[5] The output signal will result from the comparison of the channel spin polarization with the one of the FM contact.

In order the device to work, it is necessary that the dominant resistance, and hence the voltage drop, is at the interfaces, especially the one between the FM and the SGS. The energetics of the interface does matter but the literature does not present any study (theoretical and experimental) about the intimate contact between a metal and a SGS. Since the Fermi level falls in a zero gap and the Fermi level alignment is usually assumed as the thermodynamic equilibrium condition, we should not have any injection barrier at the interface. Nevertheless, when a metal is in direct contact with e semiconductor, especially if it has high resistivity, an injection barrier exists at the interface.[15] Moreover when a ferromagnetic metal and an oxide are in contact, the interface appears "dead", probably because of a partial oxidation of the metal.[16] However, a barrier is required at the FM/SGS interface also for magnetically decouple the two materials. This point is sometimes neglected in literature,[17] but it is proven to be compulsory.[18,19] We estimate such an interfacial barrier to be high as the differences in materials work functions $\phi_w$ as in the usual Schottky-Mott model. Therefore, the device is similar to a Schottky barrier MOSFET (SBMOSFET).[20–24] We adopt Aluminum ($\phi_w$ = 4.1 eV) for the non magnetic metal, Nickel ($\phi_w$ = 5.0 eV) for the ferromagnetic metal, and doped $PbPdO_2$ as SGS, considering our measured $\phi_w$ = 5.2 eV.[25] Hence the SGS work function is higher than the ones of the S and D contacts.

We also study the condition where a tunnel barrier of 10 Å and 1.5 eV is placed between FM and SGS to magnetically decouple them; these numbers are the case of amorphous $Al_2O_3$.[26]

The doped $PbPdO_2$ are $p$-type gapless semiconductors.[25,27] The gate electric field causes a Fermi level shift into the bottom part of the conduction band, forming an inversion $n$-channel with almost complete spin polarization, as discussed above. We consider a flat band voltage $V_{FB}$ of 0.1 eV (the case of gold as gate contact, $\phi_w$ = 5.1 eV, for instance).

We assume that the voltage between S and D ($V_{SD}$) drops at the S and D interfaces with SGS, where the barrier is established, and neglect the voltage drop along the SGS. We also impose the current continuity. When $V_{SD}$ is applied, one of the two interface junctions is forward biased and the other one is reverse biased. In order to maximize the voltage drop at the FM/SGS interface, we assume that this is the reverse biased junction and define it as the D. Such approach is similar to what already attempted in a somehow similar system.[28] We thus have three equations:

$$V_{SD} = V_S + V_D \tag{1}$$
$$J_{SD} = J_S \cdot (\exp(eV_S/k_BT) - 1) \tag{2}$$
$$J_{SD} = -J_D \cdot (\exp(-eV_D/k_BT) - 1) \tag{3}$$

where $V_S$ and $V_D$ are the voltage drops at S and D, $J_{SD}$ is the overall flowing current density. $J_S = A^*T^2\exp(-\phi_S/k_BT)$ and $J_D = n_sA^*T^2\exp(-\phi_D/k_BT)$, where $\phi_i$ is the barrier height at the $i$ interface, $A^*$ the effective Richardson constant,[29] $n_s$ the product of the normalized spin polarized densities of states of SGS and FM according to their parallel (P) or antiparallel (AP) configuration, as the tunnel current is proportional to this quantity.[30] In the case of the insertion of the tunnel barrier at the D contact, the multiplying factor $\exp(-\alpha_T d\sqrt{q\phi_T})$ is added in $J_D$, where $d$ is the tunneling distance (10Å), $\phi_T$ is the tunneling barrier height (1.5 eV) and $\alpha_T$ is a constant.[29]

By substituting eq. (1) in (2), and then in (3), one gets

$$J_{SD} = J_S J_D \frac{1-\exp(-eV_{SD}/k_BT)}{J_S+J_D\exp(-eV_{SD}/k_BT)} \tag{4}$$

In SBMOSFET the barrier height has a nonlinear trend with gate bias.[20,21] We assume that the change in the barrier height corresponds to half of the part of the gate bias exceeding $V_{FB}$; for instance, at D the barrier is 0.2 eV (from the work functions difference), a $V_G$ of 0.2 V makes it 0.15 eV). This last assumption is reasonably the main weakness in our model; unfortunately, we cannot solve it without experiments. About $n_s$, we assume that the normalized majority spin polarized density of states for Nickel is 0.72,[13] while for the SGS it depends on the distance between the Fermi level and the minority spin sub-band, which varies with the gate bias. For instance, for $V_G$ = 0.2 V, 0.3 V and 0.4 V, the spin polarized density of states is respectively 0.996, 0.979 and 0.854. Hence, depending on $V_G$, in the P configuration $n_s$ is 0.717, 0.704 and 0.615, while, in the AP, $n_s$ is 0.279, 0.274 and 0.239, respectively.

We describe now the behavior of eq. (4), considering the room temperature situation. Figure 2 reports the transistor characteristics for the device with Al source and Ni drain, without (a) and with (b) the Alumina tunnel barrier (TB) at the D contact. Data are plotted for two gate biases and for the parallel (P) and antiparallel (AP) configurations, as indicated. It can be observed that the output current is initially very low and increases exponentially to reach the saturation regime. The ratio between the current in the saturation region and below the threshold can be as high as $10^{13}$. The value of the current in the saturation regime can be estimated as $\lim_{V_{SD}\to\infty} J_{SD}$ and appears to be set by $J_D$, i.e. the reversed bias junction, which is reasonable. This is also the reason for the huge decrease of its value when the tunnel barrier is inserted.

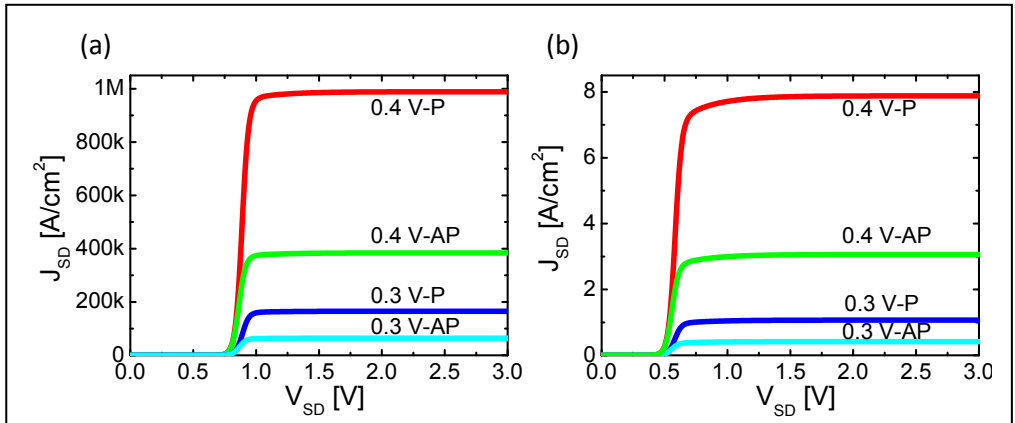

**Figure 2**: current density flowing between source and drain in respect to the source-drain voltage for the SGS based spin MOSFET without (a) and with (b) the tunnel barrier at the SGS/FM interface.

Figure 3a reports the same calculations of figure 2a but with Au ($\phi_w$ = 5.1 eV) as S metal instead of Al ($\phi_w$ = 4.1 eV). The saturation current value does not change as it depends on $J_D$, as discussed above, and, for the same reason, starts at the same $V_{SD}$. The main difference is in how the saturation regime is reached, with a low $J_{SD}$ plateau that extends up to close to 1 V in the case of Al (higher Schottky barrier) and looks not exist with Au (very low barrier). This is because most of the applied source-drain voltage drops on the reverse biased D junction. When it is much more resistive than the forward biased junction (as in the case of Au) it is more influent in determining the overall resistance. On the contrary, when the forward biased junction have higher resistance (as in the case of Al) the voltage interval with comparable voltage drop is more extended. Thus, the reverse biased junction controls the saturation regime while the resistance ratio between the two determines how it is reached. This explains also why the saturation region is reached at lower $V_{SD}$ when the TB is inserted (compare figure 2a and 2b).

The saturation current value depends more by the gate voltage than by the P/AP condition because of the exponential dependence of the current from the barrier height. Nevertheless the achievable magnetoconductance, (MC = ($J_P$ - $J_{AP}$)/$J_{AP}$) is as high as 160 % (see figure 3b) for a given gate bias, with a low dependence on the gate voltage: MC increases from about 150 % to 160 % when the gate voltage increases from 0.2 to 0.4 V. Such MC value enables this spinMOSFET for memory applications.

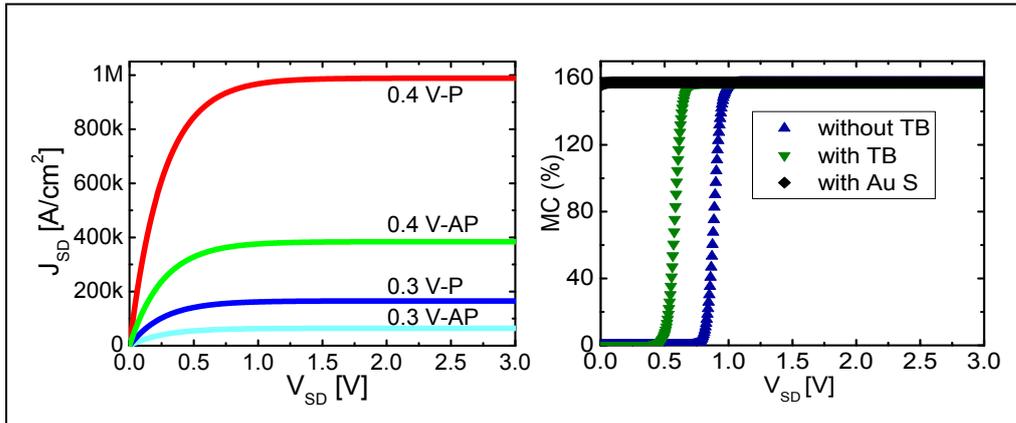

**Figure 3**: (a) current density flowing between source and drain in respect to the source-drain voltage for the SGS based spin MOSFET when Au is used as S electrode instead of Al; (b) magnetoconductance versus source-drain bias for three device configuration: the first proposed (Al source an no tunnel barrier TB) and two variations (a TB inserted at the SGS/FM interface and substitution of Al with Au).

Figure 3b reports the MC dependence on the source-drain voltage for a 0.4 V gate voltage in the three conditions of figures 2a, 2b and 3a. The MC effect is related to the forward biased contact and its value is high when this junction is dominating, i.e. when the greatest part of the voltage drops on it. Hence in the case of Al source the MC is higher in the saturation region. The insertion of the tunnel barrier (TB) increases the voltage drop at the D, then it starts to dominate at lower applied voltages. The employment of a very low barrier at S (gold) makes the forward biased D to dominate the voltage drop since the beginning, so the MC is very high also at low $V_{SD}$.

We now comment on some parameters like the transconductance $g_m$, defined as the current drivability of the gate voltage ($\partial J_{SD}/\partial V_G$), the output conductance $g_{SD} = \partial J_{SD}/\partial V_{SD}$ and the voltage gain $G_V = g_m/g_{SD}$.[5] By applying the definitions to eq. (4), it is observed that $g_m$ and $g_{DS}$ depend on $J_{DS}$, $V_G$ and $V_{SD}$, especially $g_m$ depends exponentially on $V_G$. In the two cases without the TB, $g_m$ and $g_{SD}$ are the same and are respectively about $5 \cdot 10^6$ ($2 \cdot 10^6$) A/cm$^2$V and $2.5 \cdot 10^3$ ($10^3$) A/cm$^2$V for the P (AP) condition, in the saturation region at $V_G$ = 0.4 V and $V_{DS}$ = 2 V. In the case where the TB is inserted, $g_m$ and $g_{SD}$ strongly decrease to respectively about 40 (16) A/cm$^2$V and 0.019 (0.008) A/cm$^2$V for the P (AP) condition. $G_V$ is about 2000 for all the three configurations.

The transconductance values look low, in particular with the TB, but the $G_V$ and MC values seem promising for a spin MOSFET for low power memory applications and reconfigurable logic.[5] Possible bottlenecks for the realization of such a device are mainly technological and concern the interface quality at the junctions and between the SGS and the gate oxides, including the interfacial roughness. This report shows that spin MOSFETs based on SGS appear a promising candidate for beyond CMOS spin device concepts.